\documentclass[12pt,preprint]{aastex}
\usepackage{amssymb}
\usepackage{epsfig}
\usepackage{natbib}
\usepackage{graphicx}

\newcommand       \K            {\,{\rm K}}

\newcommand       \simgt        {\gtrsim}

\newcommand       \um           {\mu{\rm m}}
\newcommand       \mum          {\,{\rm \mu m}}
\newcommand       \Teff         {T_{\rm eff}}

\newcommand       \msun         {\,{M_\odot}}
\newcommand       \Lsun         {\,{L_\odot}}

\newcommand       \simali       {\sim\,}

\countdef\decade=200
\advance\decade by \year
\countdef\hours=201
\advance\hours by \time
\divide\hours by 60
\countdef\mins=202
\advance\mins by \hours
\multiply\mins by 60
\multiply\hours by 100
\countdef\miltime=203
\advance\miltime by \hours
\advance\miltime by \time
\advance\miltime by -\mins
\def\today{\number\decade.\number\month.\number\day.\number\miltime}

\shorttitle{Crystalline Silicates in Evolved Stars}

\begin{document}

\title{
Crystalline Silicates in Evolved Stars. I.
{\it Spitzer}/IRS Spectroscopy of
IRAS\,16456-3542, 18354-0638, and 23239+5754
\\{\small DRAFT: \today ~~}
     }

\author{B.W. Jiang\altaffilmark{1},
        Ke Zhang\altaffilmark{1,2},
        Aigen Li\altaffilmark{3}, and
        C.M. Lisse\altaffilmark{4}}
\altaffiltext{1} {Department of Astronomy, Beijing Normal University,
                  Beijing 100875, China;
                  {\sf bjiang@bnu.edu.cn}}
\altaffiltext{2} {California Institute of Technology,
                  Pasadena, CA 91125, USA;
                  {\sf kzhang@caltech.edu}}
\altaffiltext{3} {Department of Physics and Astronomy,
                  University of Missouri,
                  Columbia, MO 65211, USA;
                  {\sf lia@missouri.edu}}
\altaffiltext{4} {Johns Hopkins University,
                  Applied Physics Laboratory,
                  Laurel, MD 20723, USA;
                  {\sf carey.lisse@jhuapl.edu}}

\begin{abstract}
We report the \emph{Spitzer} Infrared Spectrograph (IRS) observations
of three evolved stars: IRAS\,16456-3542, 18354-0638, and 23239+5754.
The 9.9--37.2$\mum$ \emph{Spitzer}/IRS high resolution spectra of
these three sources exhibit rich sets of enstatite-dominated crystalline
silicate emission features.
IRAS\,16456-3542 is extremely rich in crystalline silicates,
with $>$\,90\% of its silicate mass in crystalline form,
the highest to date ever reported for crystalline silicate sources.
\end{abstract}


\section{Introduction\label{sec:intro}}
The field of circumstellar silicate research started in the 1960s
when \citet{Kamijo1963PASJ...15..440K}
first theoretically investigated the condensation
of dust grains in the circumstellar envelopes around
M type long period variables and argued that {\it silica}
(SiO$_2$) should be the most abundant condensed species.
\citet{Gilman1969ApJ...155L.185G} showed several years later that grains around
oxygen-rich cool giants are mainly {\it silicates}
such as Al$_2$SiO$_3$ and Mg$_2$SiO$_4$.
The presence of silicate dust in evolved stars
was first revealed through
the detection of the 9.7$\mum$ Si--O stretching emission band
in M stars \citep{Woolf1969ApJ...155L.181W}. This was later further supported
by the detection of the 18$\mum$ O--Si--O bending emission band,
also in M stars \citep{Treffers1974ApJ...188..545T}. The 9.7 and 18$\mum$
silicate bands have also been seen in {\it absorption} in evolved stars
with an extended, heavily obscured circumstellar dust shell
(see \citealt{Jones1976ApJ...209..509J, Bedijn1987A&A...186..136B}
and references therein).

All the earlier observations showed that the 9.7 and 18$\mum$
circumstellar silicate bands were broad, smooth and featureless,
suggesting that they were predominantly amorphous
\citep{Molster2005SSRv..119....3M}.\footnote{%
  The atoms in an amorphous silicate grain are arranged
  in a random manner. In contrast, in crystalline silicates,
  the highly ordered counterparts of amorphous silicates,
  the atoms are arranged in an ordered lattice structure.
  }
Indeed, they can be well fitted with laboratory data for
amorphous olivine silicates
(e.g. see \citealt{Dorschner1995A&A...300..503D}).
This picture has dramatically changed in the 1990s,
thanks to the {\it Infrared Space Observatory} (ISO)
and the {\it Spitzer Space Telescope}.
Using the {\it Short Wavelength Spectrometer} (SWS) on board {\it ISO},
\citet{Waters1996A&A...315L.361W}
obtained the 12--45$\mum$ emission spectra of
six oxygen-rich evolved stars. They for the first time reported
the detection of crystalline silicates in evolved stars,
as revealed by the narrow emission features characteristic
of crystalline silicate minerals at wavelengths between 30 and 45$\mum$.

Crystalline silicates have now been observed in
all evolutionary stages of evolved stars:
red giants and supergiants, asymptotic giant branch (AGB) stars,
post-AGB stars, planetary nebulae (PNe)
and luminous blue variable (LBV) stars\footnote{%
   LBV stars are very bright, massive, short-lived,
   hypergiant variable stars,
   with a luminosity exceeding $10^6\Lsun$,
   a mass up to $\simali$150$\msun$,
   and a lifetime of only a few million years.
   LBVs are a stage in the evolution of very massive stars.
   They may evolve to Wolf-Rayet stars before exploding into supernovae.
   }
\citep{Waters1996A&A...315L.361W,
Waters1999LNP...523..381W,
Sylvester1999A&A...352..587S,
Molster2002A&A...382..222M, Molster2002A&A...382..184M,
Molster2002A&A...382..241M}.
Extragalactic crystalline silicates have also been found in
evolved stars by {\it Spitzer} in the Large Magellanic Cloud
(e.g. \citealt{Kastner2006ApJ...638L..29K, Sloan2006ApJ...638..472S,
Sloan2008ApJ...686.1056S, Zijlstra2006MNRAS.370.1961Z})
and in R\,71, a LBV star by {\it ISO}
\citep{Voors1999A&A...341L..67V}.
We note that like amorphous silicates, crystalline silicates
are usually found in {\it oxygen-rich} environments
\citep{Molster2005SSRv..119....3M}.\footnote{%
   In an oxygen-rich environment, the C/O ratio is smaller than one.
   All carbon atoms will be bound in CO and there will be some extra
   oxygen atoms left to form silicates and oxides
   \citep{Gail2010LNP...815...61G}.
   In contrast, if the circumstellar envelope around an evolved
   star is carbon-rich (i.e. C/O\,$>$\,1), all oxygen atoms
   will be bound in CO and therefore no silicates or oxides will
   be able to form; instead, one expects to form carbonaceous dust
   and SiC dust \citep{Gail2010LNP...815...61G}.
   }
However, the presence of crystalline silicates in
{\it carbon-rich} evolved stars has also been reported
\citep{waters1998A&A...331L..61W,
waters1998Natur.391..868W, Molster2001A&A...366..923M}.
In these cases the crystalline silicate grains were likely
produced in previous oxygen-rich mass-loss episodes
\citep{Molster2005SSRv..119....3M, Henning2010ARA&A..48...21H}.

The identification of circumstellar crystalline silicates
has been made based on the characteristic mid-IR bands of
crystalline silicates. The {\it ISO} SWS and LWS
({\it Long Wavelength Spectrometer}) spectra
of evolved stars in the 2.4 to 195$\mum$ wavelength region
allow one to identify at least 49 narrow bands which are
attributed to crystalline silicates, with distinct
emission bands at approximately 10, 18, 23, 28, 33, 40 and 60$\mum$
\citep{Molster2002A&A...382..222M, Molster2002A&A...382..184M}.
The widths and peak wavelengths of these bands
and their relative strengths contain useful information about
the crystalline silicate composition (e.g. the Fe/Mg ratio)
and stoichiometry (e.g. olivine ${\rm Mg_{2x}Fe_{2-2x}SiO_{4}}$
or pyroxene ${\rm Mg_{x}Fe_{1-x}SiO_{3}}$
where $0\le {\rm x} \le 1$;
see \citealt{Koike1993MNRAS.264..654K, Jaeger1998A&A...339..904J}).
The mid-IR spectra of evolved stars obtained by {\it ISO}
and {\it Spitzer} so far indicate that circumstellar
crystalline silicates all appear to be extremely Mg-rich
and Fe-poor or even Fe-free
(i.e. forsterite ${\rm Mg_{2}SiO_{4}}$
or enstatite ${\rm MgSiO_{3}}$;
see \citealt{Molster2005SSRv..119....3M, Henning2010ARA&A..48...21H}).

However, there are several outstanding problems about
crystalline silicates remain to be addressed:
First, the exact composition and stoichiometry of
crystalline silicate dust in evolved stars are
not precisely known. Second, the mechanism for making
circumstellar crystalline silicates remains debated.
Third, it is unclear whether (and how) the fraction
of silicate dust in crystalline form depends on
the mass loss rate, the temperature of the circumstellar
dust shell, and the evolutionary stage of an evolved star
\citep{Kemper2001A&A...369..132K, Molster2005SSRv..119....3M,
Wooden2008SSRv..138...75W, Henning2010ARA&A..48...21H}.
It has also been suggested that high crystallinity
($>$\,10\%) is typically associated with the presence
of a disk, rather than being correlated with the mass-loss rate
\citep{Kemper2001A&A...369..132K, Molster2005SSRv..119....3M}.
Further, it is widely believed that interstellar dust
originates from stardust produced by evolved stars.\footnote{%
  But also see \citet{Draine2009ASPC..414..453D}
  and \citet{Zhukovska2008A&A...479..453Z}
  who argued that the dust in the interstellar medium (ISM)
  is not really ``stardust'' but condensed
  in the ISM as the interstellar dust destruction rates are higher
  than the injection rates of stardust into the ISM
  \citep{Jones1994ApJ...433..797J}.
  }
It is puzzling that interstellar silicates are predominantly
amorphous \citep{Kemper2004ApJ...609..826K, Li2001ApJ...550L.213L,
Li2007MNRAS.382L..26L} while an appreciable fraction of circumstellar
silicates -- the primary source of
interstellar dust -- are crystalline.

To improve our understanding of these questions,
we plan to systematically explore the spectroscopic
properties of astronomical crystalline silicates and
their formation, evolution, and dependence on the properties
of the parent stars as well as their relevance to
interstellar silicates.
In this work we present the {\it Spitzer}/IRS spectra of
three evolved objects (IRAS\,16456-3542, 18354-0638, and 23239+5754)
all of which display rich sets of crystalline silicate emission features.
This is a part of our ongoing efforts for a systematic investigation
of crystalline silicates in evolved stars.

\section{Observation and Data Reduction \label{sec:obser}}
The data were obtained through
the \emph{Spitzer} Cycle 3 GO proposal
({\it Spitzer} Program ID\,30403; PI: Biwei Jiang).\footnote{%
  See {\sf http://adsabs.harvard.edu/abs/2006sptz.prop30403J}.
  }
Sixteen evolved objects (post-AGB stars and PNe) have been
observed through this program.
They were selected from the {\it Infrared Astronomical Satellite}
(IRAS) {\it Point Source Catalog} (PSC),
with their IRAS color indices satisfying
$\lg (F_{25}/F_{12}) > -0.2$ and
$\lg (F_{60}/F_{25}) < 0$.
This is to exclude young stellar objects (YSOs) or galaxies
and limit us to evolved stars \citep{van1988A&A...194..125V}.
In this work we focus on the three sources
(IRAS\,16456-3542, 18354-0638, and 23239+5754)
whose mid-IR spectra are characterized by rich sets
of crystalline silicate emission features.
The other (13) objects will be discussed elsewhere.
In  Table~\ref{tab:infrared} we list
their IR brightness (and thus colors)
and their {\it SIMBAD} classification.

These sources were observed using
the {\it Infrared Spectrograph} (IRS) on board \emph{Spitzer}
with the {\it Short-High} (SH) and {\it Long-High} (LH) modules
during Cycle 3 from  June 2006 to May 2007.
The data were taken in `staring mode' at two nod positions
along each IRS slit. Thanks to the anticipated high
brightness of the sources, no specific off-set observations for
measuring the background were needed. The journal of observation is
shown in Table\,\ref{tab:observation}.

We start the data processing from the {\it Spitzer Science Center's}
pipeline data set (version S16.1.0). The $irsclean$
program is used to combine unflatfielded image (droopres) and the
rogue pixels image from the same campaign for a compound mask,
afterwards this combined mask is applied to the flatfielded image
(bcd) to remove rogue and flagged pixels as much as possible.
The spectra are extracted from the cleaned image in
the {\it SPICE} software package and further analysis
is done in the {\it SMART} (version 6.2.6) package.

\begin{figure}
\begin{center}
\includegraphics[scale=0.60]{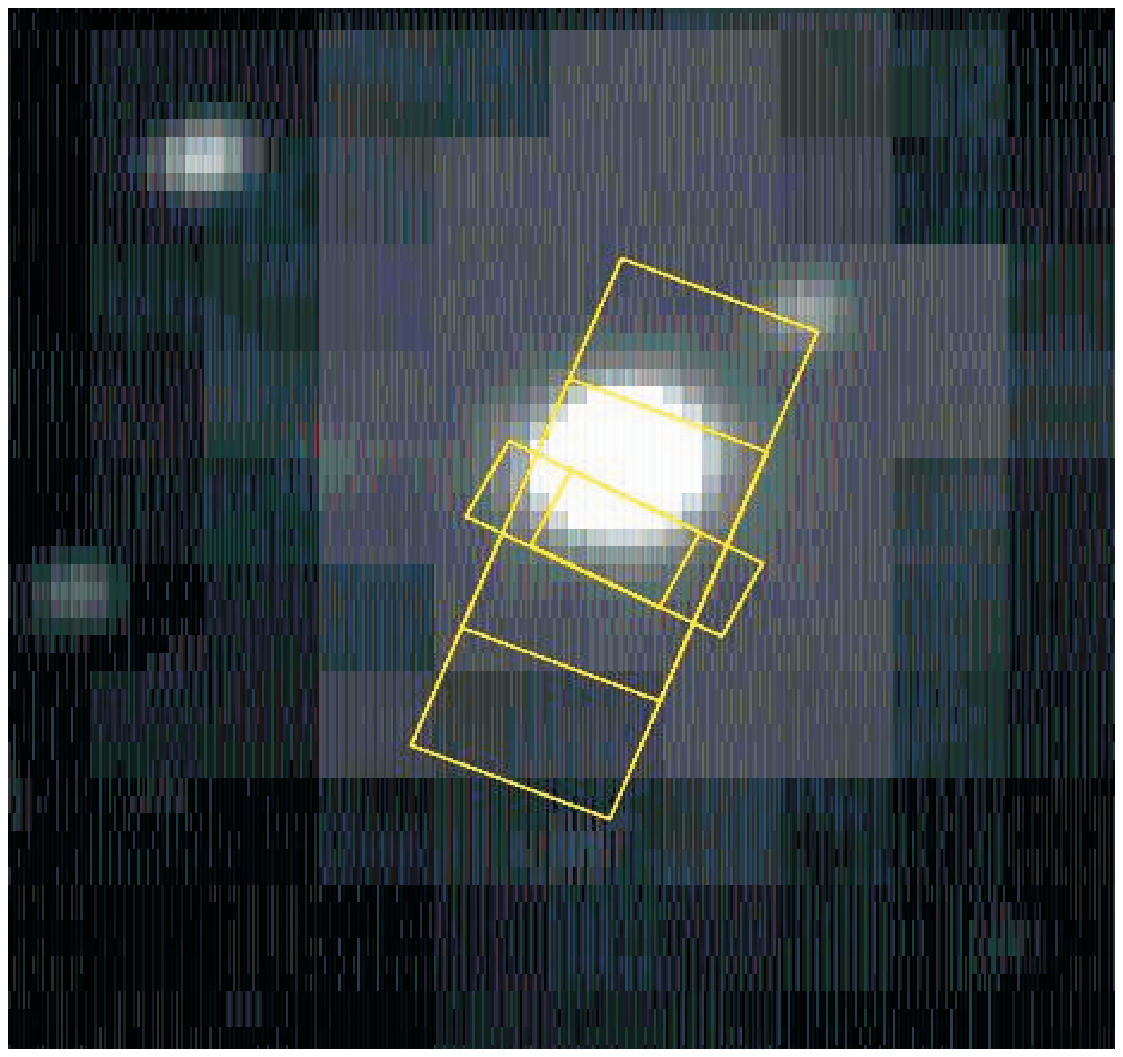}
\caption{\footnotesize \emph{Spitzer}/IRS SH (small rectangle)
and LH (large rectangle) slits superimposed on
the 2.159$\mum$ 2MASS $K$ band image (bright black and white)
and the 21.34$\mum$ MSX $E$ band image (faint extended).
\label{fig:aor} }
\end{center}
\end{figure}

Since the LH aperture (11.1$\arcsec\times$22.3$\arcsec$)
is larger than the SH aperture (4.7$\arcsec\times$11.3$\arcsec$)
and it is likely that some sources are so extended that
unlike the LH slit, the SH slit could not cover the entire
emission area, one sometimes need to multiply a scaling factor
in order for the SH flux to agree with the flux of the LH part.
The consequence of the aperture difference shows up clearly
in IRAS\,23239+5754, for which the SH spectrum must be multiplied
by a factor of $\simali$5.2 to agree with the LH flux
at the overlap part and for the whole spectrum to be consistent
with the previous {\it MSX} (Midcourse Space Experiment)
and {\it IRAS} photometry.
As shown in Figure \ref{fig:aor}, for IRAS\,23239+5754
the SH slit covers less than half of the emission range
while the LH slit covers most of the emission area,
confirming a factor of $\sim$\,5.2 difference.
The flux matching between the SH part and the LH part
is much better for IRAS\,16456-3542, and 18354-0638.
We just need a scale factor of $\simali$1.2--1.3.

\begin{figure}
\begin{center}
\includegraphics[scale=0.75,angle=-90]{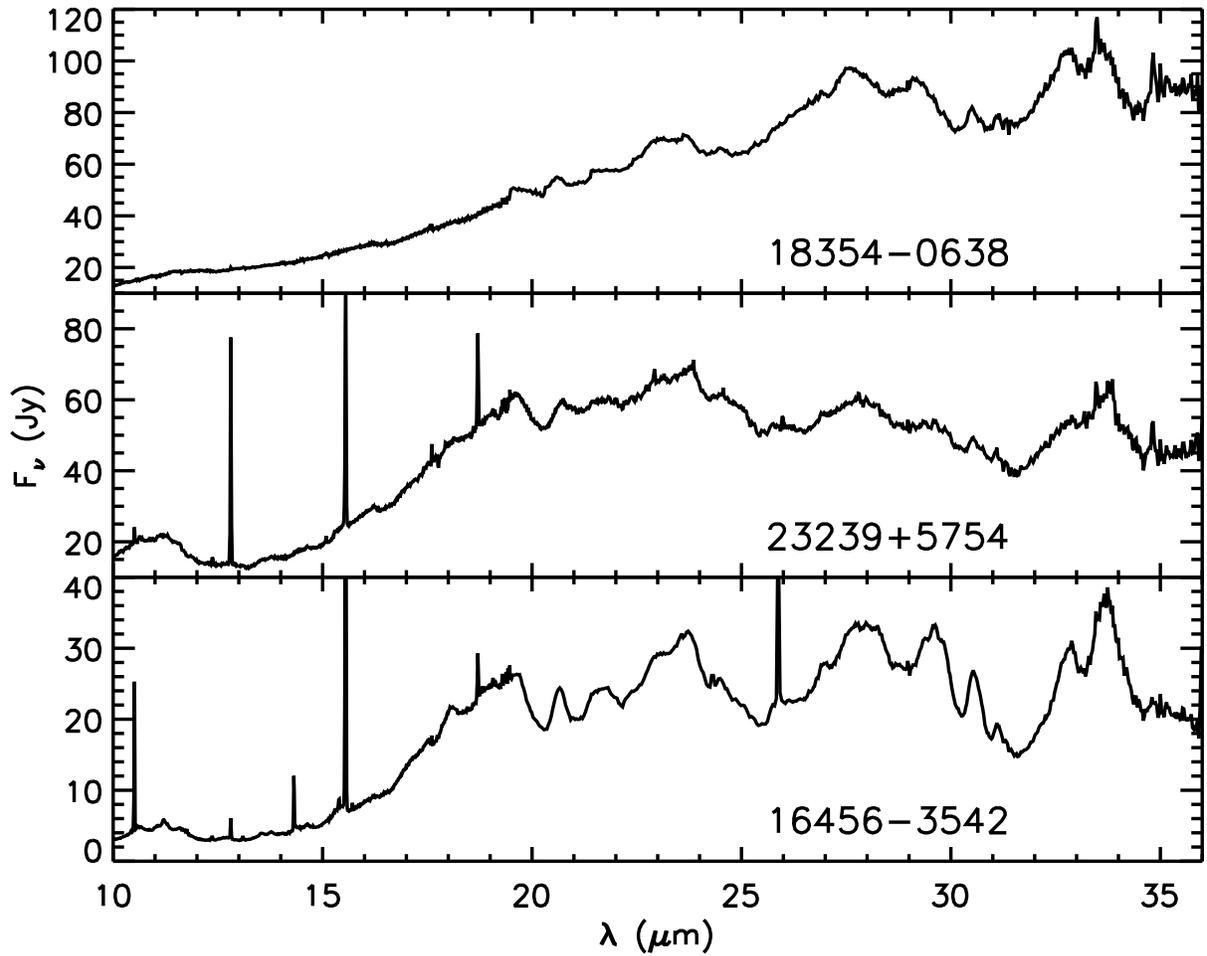}
\caption{\emph{Spitzer}/IRS spectra of
IRAS\,18354-0638, 23239+5754，and 16456-3542
(in the order of evolutionary stage from post-AGB phase
to young PN phase) which are rich in crystalline silicates.
\label{fig:spectra}}
\end{center}
\end{figure}

The reduced spectra of these three sources are displayed in
Figure \ref{fig:spectra}, where the objects are ordered in terms
of their evolutionary classes: post-AGB stars and PNe.
Based on their ionized nebular emission lines,
we classify IRAS\,16456-3542 and 23239+5754 as PNe,
consistent with the previous classifications in the literature.
For the sources lack of nebular lines, they are commonly
considered either as post-AGB stars
or as extremely evolved AGB stars.\footnote{%
  The AGB and post-AGB distinction is made based on previous studies
  available in the literature
  as the \emph{Spitzer} spectra provide no further decisive information
  for this classification.
  }
In this context we classify IRAS\,18354-0638 as a post-AGB star (see \S3.1).

\begin{figure}
\begin{center}
\includegraphics[scale=0.75]{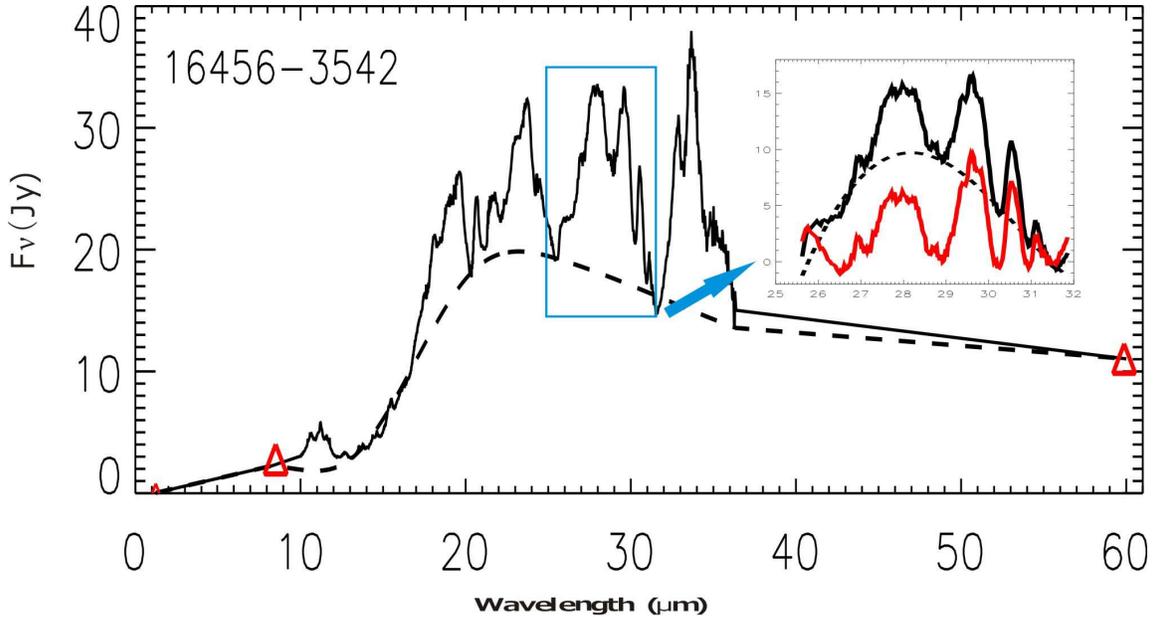}
\caption{\footnotesize Example of continuum subtraction. The
global continuum is fitted by eye-guided spline simulation (dash line)
of the observed spectrum (solid line), with the addition of three
photometry data in the 2MASS J band, MSX A band and IRAS\,60$\mum$
band denoted by red triangles. The local continuum is fitted by a
3-order polynomial with the continuum-subtracted spectral features 
shown in the inset. \label{fig:continuum}}
\end{center}
\end{figure}

%
The IRS wavelength range between 9.9$\mum$ and 37.2$\mum$
is full of spectral features arising from crystalline silicates.
In order to clearly illustrate the crystalline silicate features,
We try to subtract the underlying continuum.
Instead of adopting a single-temperature blackbody
spectrum, we simply fit the global continuum by a mathematical spline
function. This approach was taken by \citet{Molster2002A&A...382..184M}
in a comprehensive analysis of the ISO spectra of
17 oxygen-rich circumstellar dust shells
surrounding evolved stars.
For an overview of the IR
spectral energy distribution (SED),
previous photometric data,
in the {\it 2MASS} J band, {\it MSX} A band\footnote{%
  The {\it MSX } A (msxA), B$_1$ (msxB1), B$_2$ (msxB2),
  C (msxC), D (msxD),
  and E (msxE) bands respectively
  peak at 8.28, 4.29, 4.35, 12.13, 14.65, and 21.34$\mum$
  \citep{Egan1996AJ....112.2862E}.
  }
and {\it IRAS} 60$\mum$ band, are supplemented.
The continuum is then maximized and subtracted.
In principle, the strength of the features might be underestimated
since there is a possibility that very broad features are mistaken
as continuum. After this global continuum subtraction, one more step
is taken to remove the ``local continuum'' by a third-order polynomial.
The so-called ``local continuum'' refers to the part that is above the
global continuum while below the minimum of the features. This step
would possibly further underestimate the strengths of the features.
Figure \ref{fig:continuum} is an example for subtracting
the continuum underneath the spectral features.

\begin{figure}
\begin{center}
\includegraphics[scale=0.8]{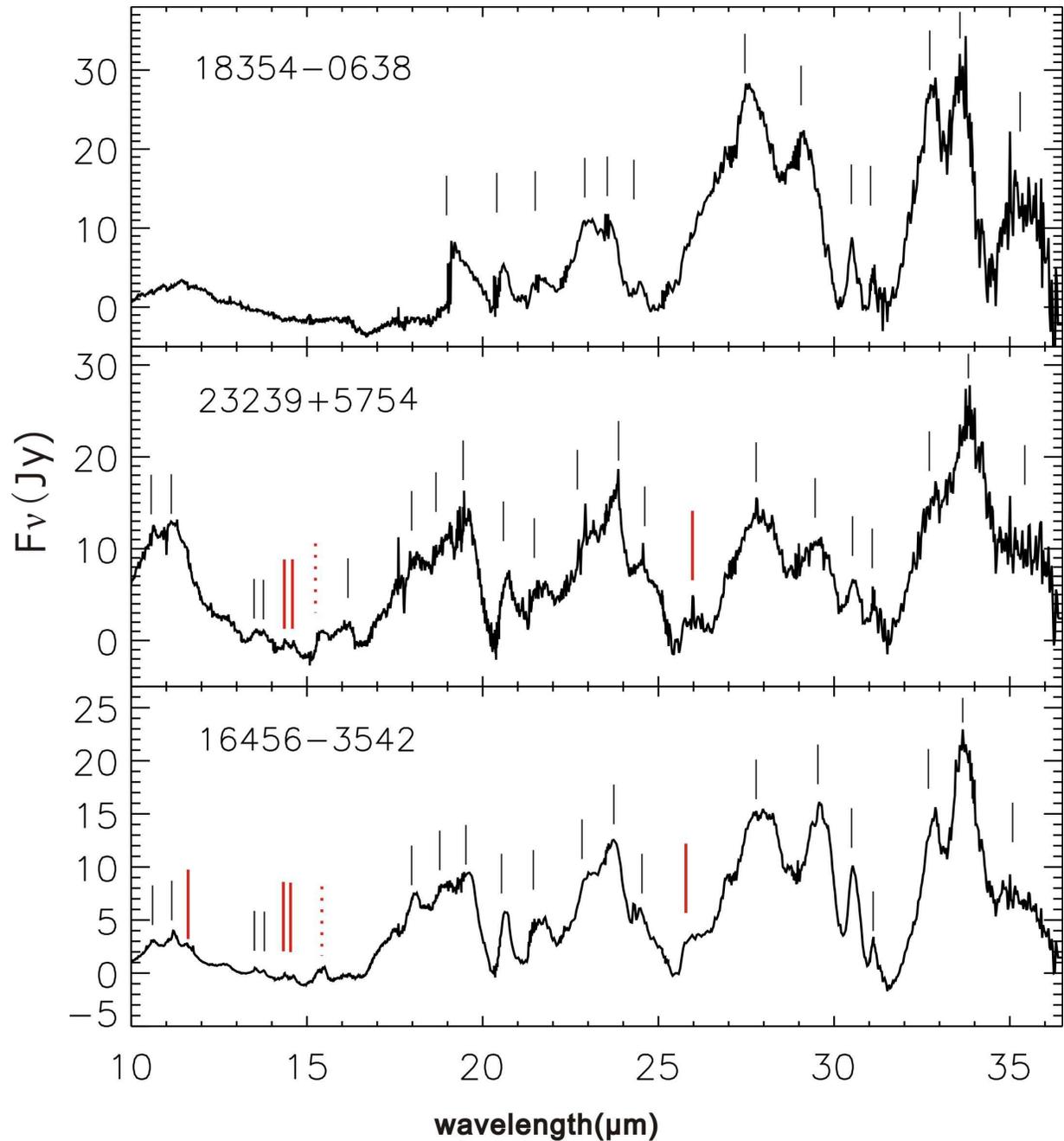}
\caption{\label{fig:features} \footnotesize
Rich crystalline silicate features in
IRAS\,16456-3542, 18354-0638, and 23239+5754.
The features already listed in \citet{Molster2002A&A...382..222M}
are labeled with thin solid black line.
The ``new features'' not reported in the literature are labeled
with thick solid red line. Those with noticeable shifts of peak
wavelength (compared to that of \citet{Molster2002A&A...382..222M})
are labeled with dotted red line. }
\end{center}
\end{figure}

Figure \ref{fig:features} displays the continuum-subtracted
spectral features of these three sources
(IRAS\,16456-3542, 18354-0638, and 23239+5754).
Table\,\ref{tab:feature} lists the peak wavelengths
and FWHM (full width half maximum) of the crystalline
silicate features derived from a Gaussian fitting.
The spectral features are compared with
the ISO spectrum of NGC\,6302, a PN
\citep{Molster2002A&A...382..184M}.
The features previously identified in the Molster sample
are labeled by thin vertical lines.

Probably due to the high resolution and
sensitivity of \emph{Spitzer}/IRS,
several ``new features''
(labeled by thick vertical lines in Figure \ref{fig:features})
appear to be present in
the IRS spectra of these sources:
(1) the 11.6$\mum$ feature seen in IRAS\,16456-3542,
(2) the 14.4$\mum$ feature (or the 14.6$\mum$ feature)
    seen in IRAS\,16456-3542 and IRAS\,23239+5754,
(3) the 15.4$\mum$ feature seen in IRAS\,23239+5754
    and in IRAS\,16456-3542,\footnote{%
      We cannot exclude the possibility that
      this feature may be the same as the 15.2$\mum$ feature
      reported in \citet{Molster2002A&A...382..184M}.
      }
    and
(4) the 25.47$\mum$ feature in IRAS\,23239+5754
    (or the 25.85$\mum$ feature in IRAS\,16456-3542).
These ``new features'' have not previously been
reported in the literature.
A more detailed study of a large sample of both astronomical
and laboratory crystalline silicate spectra is necessary for
confirming these ``new'' features.

\section{Results: Remarkably Rich Spectra of Crystalline Silicates
                  at 10--37$\mum$}
%

\subsection{Overview of the Three Objects Rich in Crystalline Silicates}
The three sources which show abundant features of
crystalline silicates are IRAS\,16456-3542, 18354-0638,
and 23239+5754. We discuss these objects below,
in the order of their evolutionary stages.

\begin{enumerate}
\item {\it IRAS\,18354-0638}.
      The \emph{Spitzer}/IRS spectrum of this source
      exhibits one nebular line (i.e. [SIII] at 33.5$\mum$)
      which may suggest the start of a nebular phase.
      We hesitate to classify it
      as a PN because PNe usually exhibit $>$\,3 nebular lines
      (while IRAS\,18354-0638 displays only one emission line).
      We classify it as a post-AGB star.\footnote{%
         We note that this star is not in the Tor\'{u}n catalog
         of post-AGB stars \citep{Szczerba2007A&A...469..799S}.
         }
      In comparison with the other two objects
      (IRAS\,23239+5754 and 16456-3542; see below)
      which display abundant crystalline silicate features
      and are classified as PNe, this object seems to be
      at an earlier stage of evolution
      (i.e. extreme AGB or post-AGB).

%

\item {\it IRAS\,23239+5754}.
      This object is a very young bipolar PN with a complex inner
      structure inside the bipolar lobes.
      The effective temperature of the central star
      is about $\Teff\approx 40,000\K$ \citep{Preite1983AAp..126...31P}.
      The mean temperature of the bulk dust in the circumstellar
      shell is $\sim$\,200\,K, as estimated from the overall
      IR spectral energy distribution
      \citep{Zhang&Kwok1990A&A...237..479Z}.
      Several nebular lines (e.g., [SIV] at 10.5$\mum$,
      [NeII] at 12.8$\mum$,
      [SIII] at 18.7$\mum$,
      [SIII] at 33.48$\mum$,
      and [NeIII] at 36.0$\mum$)
      clearly show up in its IRS spectrum.
      This source is well known for its
      pure fluorescent H$_2$ emission
      in the near IR wavelength region
      \citep{Ramsay1993MNRAS.263..695R}.
      It has long been suspected to be a binary.

      Also known as Hubble 12 at a distance of $\sim$\,3\,kpc
      \citep{Kingsburgh1992MNRAS.257..317K}, this PN,
      in particular its multiple coaxial rings in the whole-scale
      bipolar structure, has been well studied in the literature
      (see e.g. \citealt{Kwok2007ApJ...660..341K} and references therein).
      The shape of this object reveals that it is probably
      at a very early PN phase, with a kinematic age of $\simali$300 years
      \citep{Miranda1989A&A...214..353M}.
      The multiple rings extend about 4$\arcsec \times$ 8$\arcsec$.
      This size is comparable to the aperture of the IRS SH slit
      (3.7$\arcsec$\,$\times$\,11.3$\arcsec$)
      and much smaller than the IRS LH slit
      (11.1$\arcsec$\,$\times$\,22.3$\arcsec$).
      As discussed in \S2, this explains the discrepancy
      between the SH and LH fluxes.

\item {\it IRAS\,16456-3542}.
      This object is also known as PN G347.4+05.8 and H1-2.
      Its central star is a ``Weak Emission-Line Star''
      (WELS; \citealt{Gesicki2006A&A...451..925G}).
      With a dynamical age of $\simali$1400 years
      \citep{Gesicki2006A&A...451..925G},
      it is a young PN (but more evolved than IRAS\,23239+5754).
      \citet{Phillips2002ApJS..139..199P}
      derived
      a distance of $\simali$2.47\,kpc,
      but larger distances have also been suggested
      (e.g., $\simali$6.96\,kpc [\citet{Zhang1995ApJS...98..659Z}],
       $\simali$9.01\,kpc [\citet{Phillips2004MNRAS.353..589P}]).
      %
      In addition to the crystalline silicate bands,
      this object also emits at the 11.3, 12.7
      and 16.4$\mum$ PAH bands (see Figure~\ref{fig:spectra}).
\end{enumerate}


%

\subsection{Percentages of Crystalline Silicates\label{sec:model}}
To derive the mass percentage of crystalline silicates,
we follow \citet{Molster2002A&A...382..241M} to
apply a simple dust emission model to fit the observed IRS spectra
(including the continuum underlying the crystalline silicate features).
This model assumes that the dust shell is optically thin
in the IR. The SED is then fitted
with mass absorption coefficients $\kappa_{\rm abs}(\nu)$
multiplied by blackbody functions $B(T,\nu)$:
\begin{equation}\label{eq:model}
F(\nu) = \Sigma
         \left[ B(T_i,\nu)\times\kappa_{\rm abs}^{i}(\nu)
                          \times m_i \right] ~~.
\end{equation}
The dust species considered here include:
(1) amorphous silicate,\footnote{%
       For IRAS\,18354-0638 amorphous carbon fits better
       than amorphous silicates. For this object, the crystalline
       fraction is the mass fraction of crystalline silicate relative
       to the total dust mass.
       \label{ftnt:CrystFrac}
       }
(2) forsterite, and
(3) enstatite. 
For each species, we consider both a cool component
and a warm component respectively
at temperature $T_{\rm c}$ and $T_{\rm w}$.\footnote{%
   The dust temperature is treated as a free parameter
   in the range of $40\K<T<1000\K$.
   We start with a single-temperature component
   and then add a component at another temperature
   to see if the reduced $\chi^2$ significantly drops.
   }
The mass of dust species $i$, $m_{i}$, is a multiplication
factor related to the total mass and only represents the relative
mass ratio.\footnote{%
  To derive the true dust mass, the knowledge of
  the distance of the object is required.
  The relative mass ratio of different components does not
  depend on the distance and can be correctly determined
  in this way.
  }

The absorption coefficient of amorphous silicate is calculated from
the optical constants of \citet{Ossenkopf1992A&A...261..567O}
with Mie theory, assuming the dust to be spherical with radii
$a$\,=\,0.1$\mum$, the typical size of interstellar dust.\footnote{%
   The 9.7$\um$ amorphous silicate band in our sources is
   better fitted in terms of small-sized grains
   than large ones.
   }
For the minerals (forsterite and enstatite),
the optical properties measured in laboratory
depend on the size, shape, structure and chemical composition of
the dust analog.
We adopt those of \citet{koike1999} because their coefficients
closely resemble the IRS spectra of our sources.


The relative mass of each component
(amorphous silicate, forsterite and enstatite) is
treated as a free parameter. For enstatite, we consider
two sub-types: clino-enstatite and ortho-enstatite.
Following \citet{Molster2002A&A...382..241M},
we assume equal mass of ortho- and clino-enstatite.\footnote{%
   Enstatite emits at 20.65$\mum$ and 21.7$\mum$
   \citep{Molster1999A&A...350..163M}.
   The relative strengths of these two features show
   a notable difference between clino-enstatite and ortho-enstatite:
   for clino-enstatite they are about equally narrow,
   while for ortho-enstatite the 21.7$\mum$ feature is stronger
   and broader than the 20.65$\mum$ feature.
   The 20.65$\mum$ and 21.7$\mum$ features are comparably narrow
   in the three sources considered here and in AFGL\,4106,
   the source in which these two features were first identified
   by \citet{Molster1999A&A...350..163M}.
   }

We take the Levenberg-Marquart minimization algorithm to obtain
the best-fit. The results of the SED fitting are shown in
Figure \ref{fig:modelfit} and the parameters are listed
in Table\,\ref{tab:model}.
The mass percentage of crystalline silicates is
$\simali$$10\pm2\%$, $82\pm5\%$ and $97\pm2\%$
respectively for IRAS\,18354-0638 (see Footnote~\ref{ftnt:CrystFrac}),
23239+5754 and 16456-3542.
Except for IRAS\,18354-0638, the fraction of crystalline silicate
is much higher than the average of $\simali$10\% in evolved stars
\citep{Kemper2001A&A...369..132K}, and the amount of crystalline
silicate even exceeds amorphous silicate.
In IRAS\,16456-3542, the fraction of crystalline silicate reaches
$\simali$97\%, which exceeds the highest crystalline fraction
measured so far from the crystalline silicate {\it emission} spectra
($\simali$75\% in IRAS\,09425-6040
\citealt{Molster2001A&A...366..923M}).\footnote{%
   Several crystalline silicate absorption features
   were detected recently in a distant absorber
   at $z$\,$\approx$\,0.89 toward PKS\,1830-211,
   a gravitationally lensed quasar \citep{Aller2012arXiv1201.5034A}.
   The crystalline fraction derived from the {\it absorption} spectrum
   of this object may be $\simgt$95\%.
   }
Even when we take into account
the uncertainty ($<$\,5\%) in the modeling,
the fraction of crystalline silicate
in this source is one of the highest.
We conclude that there must be some evolved stars
in which the silicates are even completely crystallized.

Under the assumption that the crystalline silicate in the evolved
stars has only two forms: enstatite and forsterite, the mass ratio
of the two forms $m_{\rm e}/m_{\rm f}$ is approximately
4, 54 and 24 respectively in these three sources,
implying that the main form of crystalline silicate is
enstatite i.e. MgSiO$_{\rm 3}$. Enstatite accounts for more
than two-thirds of the crystalline silicate dust.
This result is consistent with \citet{Molster2002A&A...382..241M}
who found the enstatite to forsterite mass ratio is
between $\simali$1 and $\simali$11.4 for 12 evolved objects.

\section{Discussion\label{sec:discussion}}
\begin{figure}
\begin{center}
\includegraphics[scale=1.0]{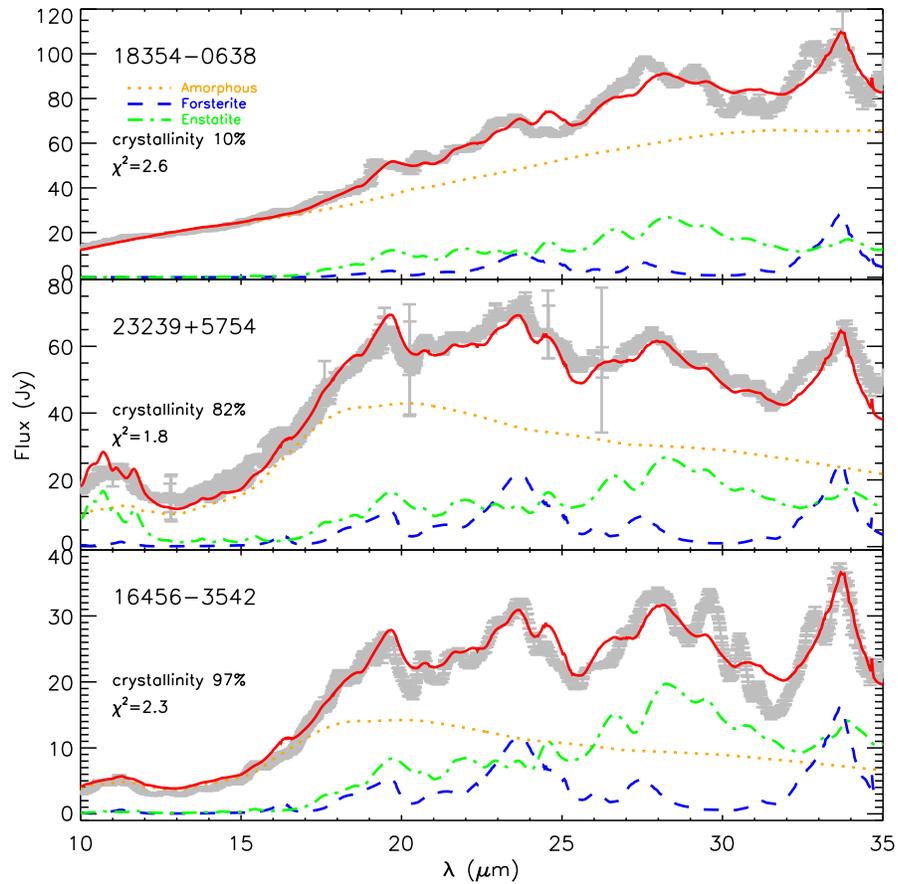}
\caption{\label{fig:modelfit}
Comparison of the model spectra (red solid line)
with the {\it Spitzer}/IRS spectra (gray solid line).
Three types of silicates are considered:
forsterite (blue dashed line),
enstatite (ortho-enstatite
and clino-enstatite; green dot-dashed line),
and amorphous silicate (red dotted line).
For IRAS\,18354-0638 we consider amorphous carbon 
instead of amorphous silicate.
\label{fig:model}}
\end{center}
\end{figure}

%
The model fit presented in \S\ref{sec:model}
shows that although most of the crystalline silicate features
can be fitted reasonably well in terms of two silicate mineral types
(i.e. forsterite and enstatite), some features are not well fitted
(see Figure \ref{fig:modelfit}).
%
%
Most noticeably, the feature at 32.7$\mum$, prominent in
all three sources and particularly strong in IRAS\,18354-0638
(in which the 32.7$\mum$ feature is nearly as strong as
the 33.5$\mum$ feature), is not accounted for by the current
model consisting of forsterite and enstatite minerals.
Similarly, the model fails in predicting the features
at 29.6, 30.6 and 31.1$\mum$ seen in all three sources
(see Figure \ref{fig:modelfit}).
We note that the 29.6, 30.6 and 31.1$\mum$
features seem to comprise a group as they have similar
relative intensities in all three sources,
with the 29.6$\mum$ feature being the strongest
and the 31.1$\mum$ feature being the weakest.
Such a phenomenon was also seen in the larger sample of
\citet{Molster2002A&A...382..222M}.
One may speculate that these features arise from
a dust mineral other than enstatite and forsterite.
It is also interesting to note that the overall
spectral features at 10--35$\mum$
of IRAS\,23239+5754 and IRAS\,16456-3542 are similar
to each other (probably because they are both very young PNe),
except that the 29.6$\mum$ feature is
much stronger in the older PN IRAS\,16456-3542
than that in the younger PN IRAS\,23239+5754.
%

\begin{figure}
\begin{center}
\includegraphics[scale=0.7]{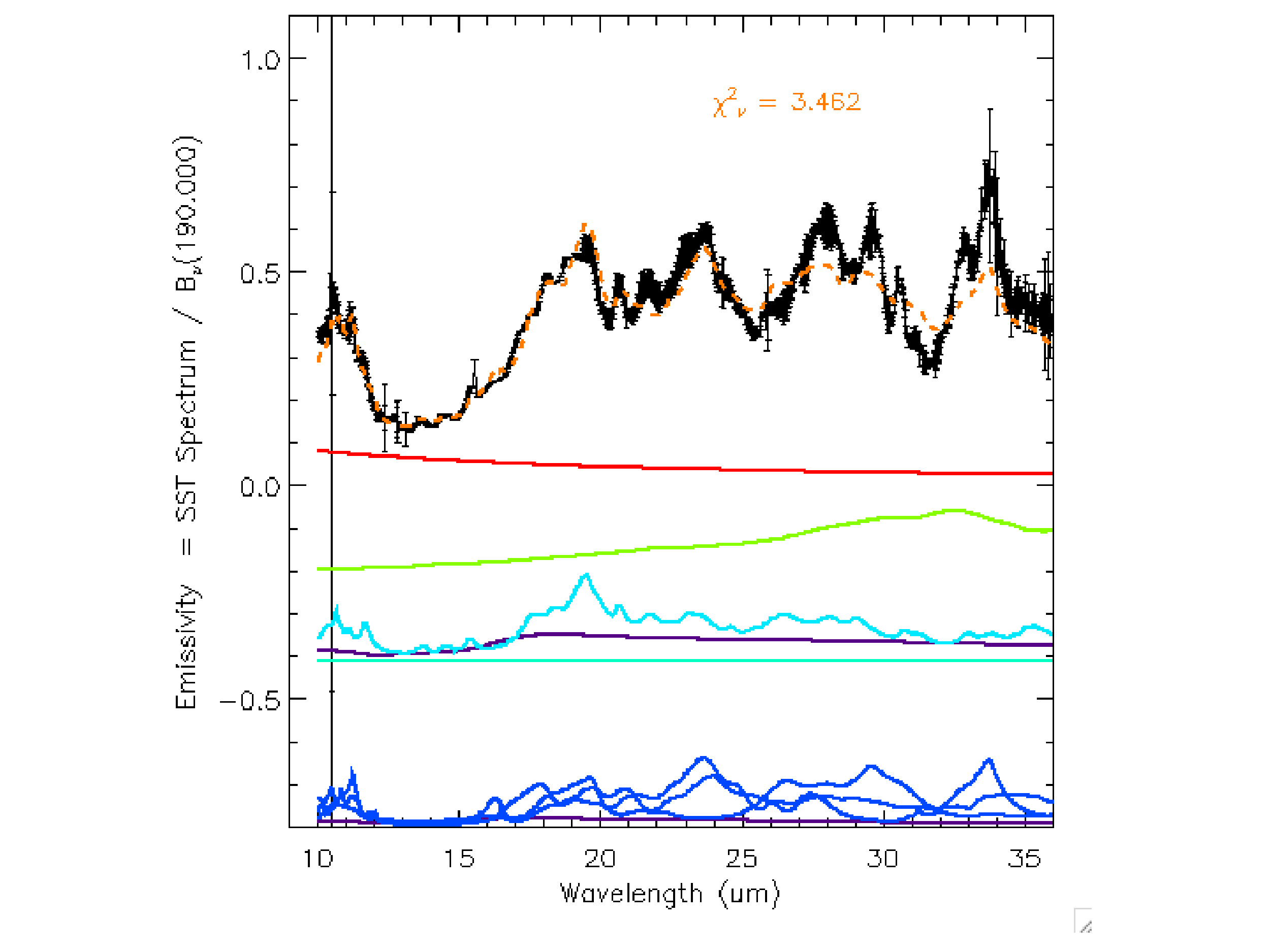}
\caption{\label{fig:LisseModel}
Modeling the {\it Spitzer}/IRS spectrum of
IRAS\,16456-3542 with the approach of \citet{Lisse07,Lisse08}
which incorporates a power-law size distribution
of $\simali$$a^{-3.9}$ with 0.1$\mum$$<$\,$a$\,$<$\,10$\mum$
for the grain size
and various crystalline and amorphous olivine and pyroxene grain
materials. The solid lines beneath the IRS spectrum show three
components: the upper red and green lines for amorphous carbon
and carbonates, the middle cyan lines for pyroxenes and sulfides,
the lower blue lines for olivines.
}
\end{center}
\end{figure}

In summary, the modeling approach discussed in \S\ref{sec:model}
assumes three grain types:
(1) spherical amorphous silicates of radii $a$\,=\,0.1$\mum$
with the indices of refraction taken from \citet{Ossenkopf1992A&A...261..567O};
(2) crystalline forsterite and
(3) crystalline enstatite for which the mass absorption
coefficients are taken from \citet{koike1999}.\footnote{%
   For crystalline forsterite and enstatite we do not need to
   assume grain size, shape, etc.
   }
For each grain type, two temperatures are assumed
(i.e. a cold component and a warm component) unless a single
temperature model results in more or less the same $\chi^2$.
To examine whether the silicate crystallinity
derived in \S\ref{sec:model} is dependent on the modeling technique,
we take a different approach:
we follow the approach documented in detail
by \citet{Lisse07, Lisse08} and incorporate
more sizes and more silicate mineral compositions. 
The detailed mineralogy of the dust is determined
by comparing the observed {\it Spitzer}/IRS spectrum
to a wide range of plausible silicate minerals with high-quality
mid-IR transmission and thermal emission laboratory spectra.
The calculated model flux from each mineralogical component
is the product of a blackbody of characteristic temperature
times the experimentally determined absorption efficiency for
each sized particle, summed over all particles in the dust population.
For illustration, in Figure \ref{fig:LisseModel} we show
the {\it Spitzer}/IRS spectrum of IRAS\,16456-3542
and the model spectrum as well as the appropriately scaled
contributions from each grain species.
%
The overall fit is reasonably good, although not as good as that
described in \S\ref{sec:model} (see Figure~\ref{fig:modelfit}).
The derived silicate crystallinity is $\simali$90\%,
consistent with the silicate crystallinity of
$\simali$97\% derived in \S\ref{sec:model}.

Finally, we note that various techniques have been proposed
in the literature to fit the crystalline silicate emission spectra
(e.g., \citealt{Molster2002A&A...382..222M, Honda2004ApJ...610L..49H, 
Sargent2006ApJ...645..395S, Lisse07, Gielen08}).
All these approaches treat dust size, composition and temperature
as free parameters. This is physically {\it incorrect} (see
\citealt{Li2004ApJ...613L.145L}).
The equilibrium temperature of dust is sensitive to its size and composition:
smaller grains have higher temperatures, and silicates
with a higher Fe/Mg ratio have higher temperatures.
The dust temperature is also sensitive to the distance
from the dust to the illuminating source (see \citealt{Li2004ApJ...613L.145L}).
Therefore, one cannot treat the dust temperature as an adjustable
parameter independent of the dust composition and size.
The dust mass is very sensitive to the dust temperature.
The crystalline fraction derived from all the approaches in which the dust
temperature is treated as a free parameter should be taken with caution.
The correct approach should physically calculate
the dust temperature for a given dust size of a given composition
at a given distance from the illuminating source
(e.g., see \citealt{Li1998A&A...331..291L, Chen2007ApJ...666..466C}).

\section{Summary \label{sec:summary}}
We report the 9.9--37.2$\mum$ spectra of three evolved objects
(post-AGB stars and PNe) using the high-resolution mode of
the {\it IRS} instrument on board {\it Spitzer}.
%
%
These objects exhibit rich sets of
      crystalline silicate emission features.
      These features are fitted by a mixture of
      three dust species: amorphous silicate,
      crystalline forsterite, and crystalline enstatite,
      with each species consisting of a warm component
      and a cold component.
      It is found that the crystalline silicates are
      dominated by crystalline enstatite.
%
The crystalline silicate spectral modeling
      shows that IRAS\,16456-3542,
      has a degree of silicate crystallinity
      (i.e. mass fraction of silicate dust in crystalline form)
      of $\simali$97\%,
      the highest to date in crystalline silicate sources.


%

\acknowledgments{We thank the anonymous referee for his/her
very constructive comments and suggestions which substantially
improved the quality of this paper. We thank B.A. Sargent
and R. Szczerba for helpful discussion.
We thank C.A. Beichman, C. Kemper, and F. Molster
for providing the IR spectra of HD\,69830, HD\,100546,
comet Hale-Bopp, and the Galactic center source Sgr A$^\ast$.
This project is supported through a NASA Spitzer GO grant,
and in part by China's grants NSFC 10973004
and 11173007 and by NSF/AST 1109039.}


\bibliographystyle{apj}



\begin{deluxetable}{lccccccccccc}
\tablecolumns{12} \tablecaption{Infrared properties}

\tablehead{ \colhead{IRAS\,Name}&\colhead{Type}& \multicolumn{3}{c}{2MASS (magnitude)} & \multicolumn{4}{c}{MSX (Jy)} &
\multicolumn{3}{c}{IRAS\,(Jy)} \\ \hline
\colhead{$$} & \colhead{$$} & \colhead{$J$}& \colhead{$H$} & \colhead{$K$}& \colhead{$A$}& \colhead{$C$}& \colhead{$D$} & \colhead{$E$}&
\colhead{$12\mum$}& \colhead{$25\mum$}& \colhead{$60\mum$} } \tabletypesize{\scriptsize} \startdata
18354$-$0638    &   Post-AGB   &   8.55    &   5.85    &   4.51    &   9.10    &   19.60   &   24.40   &   49.10   &   18.90   &   68.00   &   39.20   \\
23239+5754  &   Young PN    &   10.23   &   9.83    &   8.81    &   10.99   &   16.95   &   18.53   &   50.81   &   20.86   &   71.22   &   35.79   \\
16456$-$3542    &   PN    &   11.83   &   11.48   &   10.53   &   2.26    &   3.84    &   7.06    &   22.19   &   4.88    &   29.61   &   10.96   \\
\enddata
\label{tab:infrared}
\end{deluxetable}

\begin{deluxetable}{clcccc}
\tablecolumns{6} \tablewidth{0pc}

\tablecaption{Journal of the \emph{Spitzer}/IRS observations}

\tablehead{ \colhead{Star No.}   & \colhead{IRAS\,Name}& \colhead{MSX6C Name} & \colhead{Date observed} & \colhead{Data Version}&
\colhead{Comment}} \tabletypesize{\scriptsize} \startdata
1   &   18354$-$0638    &   G025.5470$-$00.0529 &   04/20/07    &   S16.1.0 &       \\
2   &   23239+5754  &   G111.8777$-$02.8509 &   09/09/06    &   partial S16.1.0    &   poor LH nod2    \\
3  &   16456$-$3542    &   G355.7334$-$03.4721 &   03/25/07    &   partial S16.1.0 &       \\
\enddata
\label{tab:observation}

\end{deluxetable}

\begin{deluxetable}{cc|cc|cc}
\tablecolumns{6}  \tablewidth{0pc}

\tablecaption{Inventory of crystalline silicate features of IRAS\,16456-3542, 18354-0638, and 23239+5754 \label{tab:feature}}

\tablehead{\multicolumn{2}{c|}{IRAS\,18354-0638} & \multicolumn{2}{c|}{IRAS\,23239+5754} & \multicolumn{2}{c}{IRAS\,16456-3542}\\ \hline
\colhead{${\rm \lambda_{peak}}$ } & \colhead{FWHM} & \colhead{${\rm \lambda_{peak}}$ } & \colhead{FWHM}  & \colhead{${\rm \lambda_{peak}}$ } &
\colhead{FWHM}
 }
\tabletypesize{\footnotesize} \startdata
     &      & 10.63 & 0.11 & 10.60&0.24\\
     &      & 11.26 & 0.39 & 11.22&0.17\\
     &      &       &      & \textbf{11.61}&\textbf{0.27}\\
     &      & 13.54 & 0.22 & 13.55&0.25\\
     &      & 13.75 & 0.25 & 13.75&0.23\\
     &      & 14.18 & 0.12 &      &    \\
     &      & \textbf{14.40}& \textbf{0.13} &\textbf{14.39}&\textbf{0.26}\\
     &      & \textbf{14.63} & \textbf{0.08} & \textbf{14.61}&\textbf{0.19}\\
     &      & \textit{15.45} & \textit{0.33} & 1\textit{5.39}&\textit{0.31}\\
     &      & 16.10 & 0.60 & 16.10&0.30\\
     &      & 17.47 & 0.99 & 17.43&0.39\\
     &      & 18.00 & 0.53 & 18.96&0.42\\
19.45& 0.83 & 19.59 & 0.54 & 19.61&0.50\\
20.65& 0.34 & 20.70 & 0.31 & 20.65&0.29\\
21.66& 0.47 & 21.60 & 0.45 & 21.60&0.53\\
     &      & 23.09 & 0.54 & 23.07&0.50\\
23.21& 1.21 & 23.74 & 0.49 & 23.68&0.53\\
24.49& 0.23 & 24.58 & 0.39 & 24.52&0.24\\
     &      & \textbf{25.47} & \textbf{0.55} & \textbf{25.85}&\textbf{0.30}\\
27.73& 0.85 & 27.72 & 1.30 & 27.88&1.14\\
29.19& 0.85 & 29.53 & 0.86 & 29.58&0.66\\
30.54& 0.29 & 30.59 & 0.36 & 30.57&0.27\\
31.11& 0.19 & 31.12 & 0.14 & 31.13&0.14\\
32.78& 0.95 & 32.81 & 0.50 & 32.80&0.42\\
33.57& 0.88 & 33.80 & 0.93 & 33.72&0.62\\
35.41& 1.36 & 35.64 & 1.55 & 35.55&1.14\\
\enddata
\tablecomments{Compared with the crystalline silicate feature list of
  \citet{Molster2002A&A...382..222M}, the features in Roman are at the
  same wavelength as that of \citet{Molster2002A&A...382..222M}. The
  features peaking at appreciably different wavelengths are listed in
  italic. Those listed in bold face are new features, not identified previously.}
\end{deluxetable}

\begin{deluxetable}{ccccccccc}
\tablecolumns{9}  \tablecaption{Dust parameters derived from crystalline silicate spectral modeling (see \S\ref{sec:model}) \label{tab:model}} \tablehead{Source& $T_{\rm cf}$
&$T_{\rm wf}$ &$T_{\rm ce}$ &$T_{\rm we}$ &$T_{\rm ca}$ &$T_{\rm wa}$ & $m_{\rm e}/m_{\rm f}$ & Crystallinity(\%)} \startdata
     18354$-$0638 &     79 & $\cdots$      &    87 &   $\cdots$      &    85 &   213 &   4 &   $10\pm2$\\
     23239$+$5754 &      $\cdots$    &   129 &    81 &    $\cdots$     &    $\cdots$     &   150 &  54 &   $82\pm5$\\
      16456$-$3542 &     67 &   125 &    72 &   144 &     $\cdots$    &   158 &  24 &   $97\pm2$\\
\enddata
\tablecomments{From left to right: the source, the temperatures of the
  cool forsterite component ($T_{\rm cf}$), the warm forsterite
  component ($T_{\rm wf}$), the cool enstatite component ($T_{\rm
    ce}$), the warm enstatite component ($T_{\rm we}$), the cool
  amorphous silicate component ($T_{\rm ca}$), and the warm amorphous
  silicate component ($T_{\rm wa}$), the mass ratio of enstatite to
  forsterite ($m_{\rm e}/m_{\rm f}$), and the derived silicate
  crystallinity (i.e. the mass fraction of silicate dust in crystalline form).}
\end{deluxetable}


\end{document}